\documentclass[%
 reprint,
 amsmath,amssymb,
 aps,
pra,
]{revtex4-2}

\usepackage{graphicx}
\usepackage{dcolumn}
\usepackage{bm}
\usepackage[utf8]{inputenc}
\usepackage[T1]{fontenc}
\usepackage{xcolor}
\usepackage{soul}
\usepackage{ragged2e}  
\usepackage{titlesec}  
\usepackage{hyperref}

\DeclareUnicodeCharacter{0308}{\"{e}}

\titleformat{\section}  
  {\raggedright\normalfont\bfseries\normalsize}  
  {\thesection}{1em}{}  

\titleformat{\subsection}  
  {\raggedright\normalfont\bfseries\normalsize}  
  {\thesubsection}{1em}{}  
\renewcommand*{\thesection}{\arabic{section}}
\renewcommand*{\thesubsection}{\thesection.\arabic{subsection}}

\begin{document}

\preprint{APS/123-QED}

\title{Nematic ordering via vertical stratification in drying clay nanotube suspensions}

\author{Arun Dadwal$^{1}$}
\author{Meenu Prasher$^{2}$}
\author{Nitin Kumar$^{1}$}
\email{nkumar@iitb.ac.in}

\affiliation{
	$^1$Department of Physics, Indian Institute of Technology Bombay Powai, Mumbai 400076, India. \\
 $^2$Materials Science Division, Bhabha Atomic Research Centre, Mumbai 400085, India 
	}
\date{\today}






\begin{abstract}
Evaporative self-assembly offers a simple, cost-effective method for producing functional nanostructured materials. However, achieving tunable and ordered assemblies remains challenging, especially when working with complex building blocks like nanoparticles that exhibit significant shape and size polydispersity. In this study, starting from an aqueous suspension of a polydisperse sample of rod-like Halloysite nanotubes, we present a physical protocol for producing a high degree of orientational ordering in the final dried deposit. By placing a sessile droplet on a substrate heated to 50$^\circ $C, self-induced Marangoni flows suppress the coffee-ring effect, enabling more uniform deposition of colloidal rods. Subsequently, the vertical stratification during evaporation leads to the segregation of particles by aspect ratio, with longer rods (aspect ratio $\geq$ 6.5) preferentially migrating to the top layers over the entire deposit. Since rods exceeding this threshold exhibit nematic ordering at high densities, the resulting top layer, spanning an area of the order of mm$^2$, displays a high degree of orientational order. Thus, our results highlight a robust strategy for engineering ordered structures from disordered colloidal suspensions despite the overall polydispersity of the system.

\end{abstract}

\maketitle


\section{Introduction}
The process of colloidal self-assembly due to evaporation has sparked significant interest in scientists and engineers \cite{byun2010hierarchically,han2012learning,li2013macroscopic,li2016evaporative,ren2021dendritic, daware2024synthesis,li2024evaporative}. The earliest systematic investigations into this phenomenon started with the classic observation where solute particles suspended within a sessile droplet migrate to the periphery during evaporation, leaving a characteristic ring-shaped stain, known as the \textit{coffee-ring} effect \cite{deegan1997capillary,deegan2000pattern}. This effect, though initially seen as a limitation in achieving uniform coatings, led to a series of studies highlighting the role of underlying hydrodynamic flows, particularly the role of capillary and Marangoni forces \cite{hu2006marangoni, patil2016effects, gelderblom2022evaporation}. These insights eventually enabled strategies to suppress ring formation and achieve more homogeneous particle deposition \cite{li2016rate, lama2020modulation}.

Most research in this area has rightly focused on spherical nanoparticles at low-volume fractions, aiming primarily to uncover the fundamental physical mechanisms at play. To generalize the principles of evaporative self-assembly to real-world complex fluids that are inherently anisotropic and polydisperse, it is essential to first develop simplified model systems. One such potential system can be a collection of anisotropic particles, such as rods, that are known to exhibit a rich liquid crystalline phases like nematic and smectic phases \cite{song2003nematic,ming2008ordered,nobile2009self, dugyala2015evaporation,dugyala2015evaporation, zhao2015orientation,paineau2016liquid,khawas2023anisotropic,dadwal2023quantifying,vaisakh2025role,zhang2018interplay,tan20212d,kumar2022catapulting,almohammadi2023evaporation,sharma2024phases}. Yet, studies addressing how polydispersity in their aspect ratio affects this ordering during evaporation are scarce. Given the potential relevance of such two-dimensional (2D) ordered assemblies in various technological \cite{shi2020single} and biomedical applications \cite{sefiane2010formation, bhatt2024front},  this demands further scientific investigation and experimental exploration. In this paper, we take a small step in this direction and address one particular engineering challenge: can we create a highly nematic, uniform deposition of rod-like nanotubes, starting from a rather polydisperse sample?

To address this challenge, we propose a method that leverages the intriguing phenomenon of vertical stratification during evaporation, originally introduced in the seminal work of Routh and Zimmerman in \cite{routh2004distribution} and further developed in subsequent studies \cite{trueman2012autostratification,trueman2012auto,fortini2016dynamic,zhou2017cross,howard2017stratification,liu2018stratification,schulz2018critical,liu2019sandwich,carr2024experimental,hooiveld2025self}. These studies established that when the evaporation rate dominates over thermal diffusion, particles of varying sizes segregate vertically within a drying film. Larger particles, owing to their lower diffusivity, remain near the air-water interface longer, resulting in their accumulation at the top \cite{nikiforow2010self}. Although these analyses primarily considered flat films, later theoretical and experimental work showed that similar stratification effects also occur in evaporating sessile droplets \cite{cusola2018particulate}. However, to the best of our knowledge, experimental studies investigating the stratification of rod-shaped colloids are currently lacking in the field. Additionally, we integrate vertical stratification with the fact that rod-like particles undergo an isotropic-to-nematic transition beyond a critical aspect ratio of $\approx 7$ in lyotropic nematogens \cite{Bates2000,ghosh2007orientational}. Thus, our approach leverages the combined effects of geometric anisotropy and evaporation-driven stratification to produce highly nematic and spatially uniform assemblies, even when starting from a polydisperse colloidal suspension.

To this end,  we introduce our model system, which is a colloidal suspension of Halloysite Nanotubes (HNTs), also referred to as rods in this paper, a naturally occurring clay mineral characterized by a highly polydisperse aspect ratio \cite{lvov2008halloysite,massaro2017halloysite,lisuzzo2019colloidal}. We perform controlled drying experiments using its charge-stabilized suspensions at a 5 wt.$\%$ concentration on a substrate heated to 50$^\circ$C. Using Scanning Electron Microscopy (SEM), we show that the entire top layer of the resulting deposit exhibits a high degree of orientational order, quantified using the nematic order parameter. Cross-sectional SEM imaging confirms that the enhanced nematic order arises from vertical stratification, i.e., longer rods predominantly occupy the top layers, while shorter ones accumulate near the substrate. In contrast, when the same suspension is dried on an unheated substrate at room temperature, nematic ordering is restricted to a narrow region near the coffee-ring, while the central region remains largely disordered, with no signatures of vertical stratification. Interestingly, heating the substrate to an even higher temperature of 80$^\circ$C leads to a reduction in nematic order, driven by the emergence of topological defects. This suggests the existence of an optimal substrate temperature to observe this phenomenon, which is close to 50$^\circ$C in our experiments. Thus, our protocol exploits a delicate interplay between evaporation-driven flows, vertical stratification, and the intrinsic tendency of longer rods to undergo nematic ordering across the entire dried deposit. Finally, we place our findings within the broader framework of \text{Pé}clet number–driven vertical stratification, drawing comparisons with previous studies and underscoring the distinct features and new insights uncovered in our work.


\begin{figure}[htbp]
	\centerline{\includegraphics[width=0.45\textwidth]{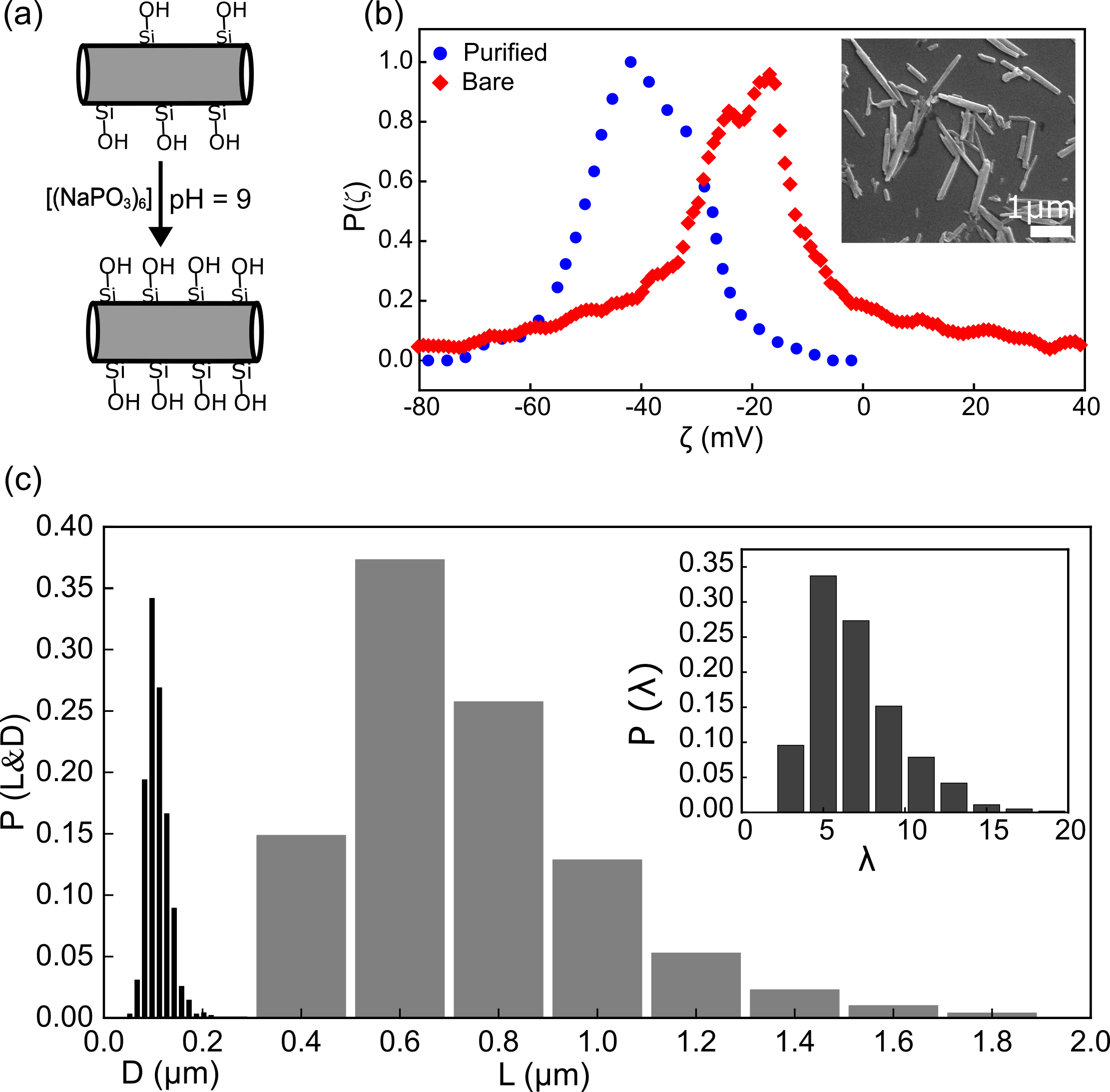}}
	\caption{Physicochemical characterization of purified HNTs: (a) Schematic illustration of the HNTs purification process that involves treating them with NaOH to increase surface group density of Si–OH via mild etching, while SHMP improves dispersion by electrostatically interacting with the positively charged lumen. (b) Zeta potential comparison of bare HNTs with purified HNTs at pH = 9, with the inset showing an SEM image of the purified HNTs. (c) Length ($L$) and diameter ($D$) distributions of the purified HNTs, with the inset showing the aspect ratio distribution, $P(\lambda)$, highlighting the polydisperse nature of the sample.}
	\label{Figure-1}
\end{figure}
\section{Results \texorpdfstring{$\textbf{\&}$}{\&} Discussion}
\subsection{Purification and Characterization of HNTs}

Pristine HNTs were purified following the procedure presented in the methods Section 4.1. Briefly, this process involves treating them with sodium hexametaphosphate [(NaPO$_{3}$)$_{6}$](SHMP) under alkaline conditions, as illustrated in Figure 1a. SHMP serves as an effective dispersing agent by enhancing the electrostatic repulsion between HNTs, thereby minimizing their aggregation \cite{xu2024formation}.  Upon addition of SHMP, the ionized phosphate groups (PO$_4^{3-}$) can penetrate the positively charged lumen and neutralize the internal charge via electrostatic interactions. Furthermore, treatment with NaOH at pH $\approx$ 9 under continuous stirring for 24 hours can induce partial etching of the HNT walls, resulting in an increased density of surface silanol (Si--OH) groups \cite{xu2024formation, zeng2014facile}. The effectiveness of this treatment is demonstrated in Figure 1b, which shows the zeta potential ($\zeta$) of untreated and SHMP-treated HNTs at pH 9. The bare HNTs exhibit a $\zeta$ value of \textendash16.7\,\text{mV}, while the treated sample shows a significantly higher negative value of \textendash42\,\text{mV}, confirming the enhancement of the surface charge. 

To quantify the extent of polydispersity, we analyze multiple SEM images where the HNTs were well-dispersed, manually measuring both their lengths ($L$) and diameters ($D$) using ImageJ software \cite{abramoff2004image}. A representative SEM image of the purified HNTs is shown in the inset of Figure 1b. The resulting distributions are presented as histograms in Figure 1c. The average and standard deviation of the lengths and diameters are 0.74 $\pm$ 0.25 $\mu$m and 0.11 $\pm$ 0.02 $\mu$m, respectively. Notably, the relative width [standard deviation (SD) divided by mean] is significantly higher for $L$ ($= 0.34$) than for $D$ ($= 0.18$), indicating that most of the polydispersity arises from length variations in our system. Defining their aspect ratio as $\lambda = L/D $, and analyzing approximately $ 1000$ rods, we plot its distribution in the inset of Figure 1c, showing a mean and SD of 7.0 and 2.8, respectively. These measurements confirm that our system is highly polydisperse, primarily due to the broad distribution in HNT lengths.
\subsection{Experimental Procedure}
Figure 2a illustrates the schematic of the droplet-drying experimental setup. In each experiment, a sessile droplet of volume 4 $\mu$L, containing HNTs at a concentration of 5 wt.$\%$, is placed onto a silicon substrate heated to 50$^\circ$C. While the relative humidity is not actively controlled, it remains stable in the range of 60 $\pm$5 $\%$ in all our experiments. For comparison, control experiments are also performed at room temperature on unheated substrates. To ensure consistency, all substrates are cleaned prior to deposition following the procedure described in Section 4.2 of the methods, such that the initial contact angle ($\theta$) of the droplet is always $50^\circ \pm 5^\circ$ in both heated and unheated cases. A relatively high contact angle is chosen to allow a sufficient vertical gap, which is expected to facilitate the vertical stratification of rods during evaporation.

\begin{figure}
	\centerline{\includegraphics[width=0.48\textwidth]{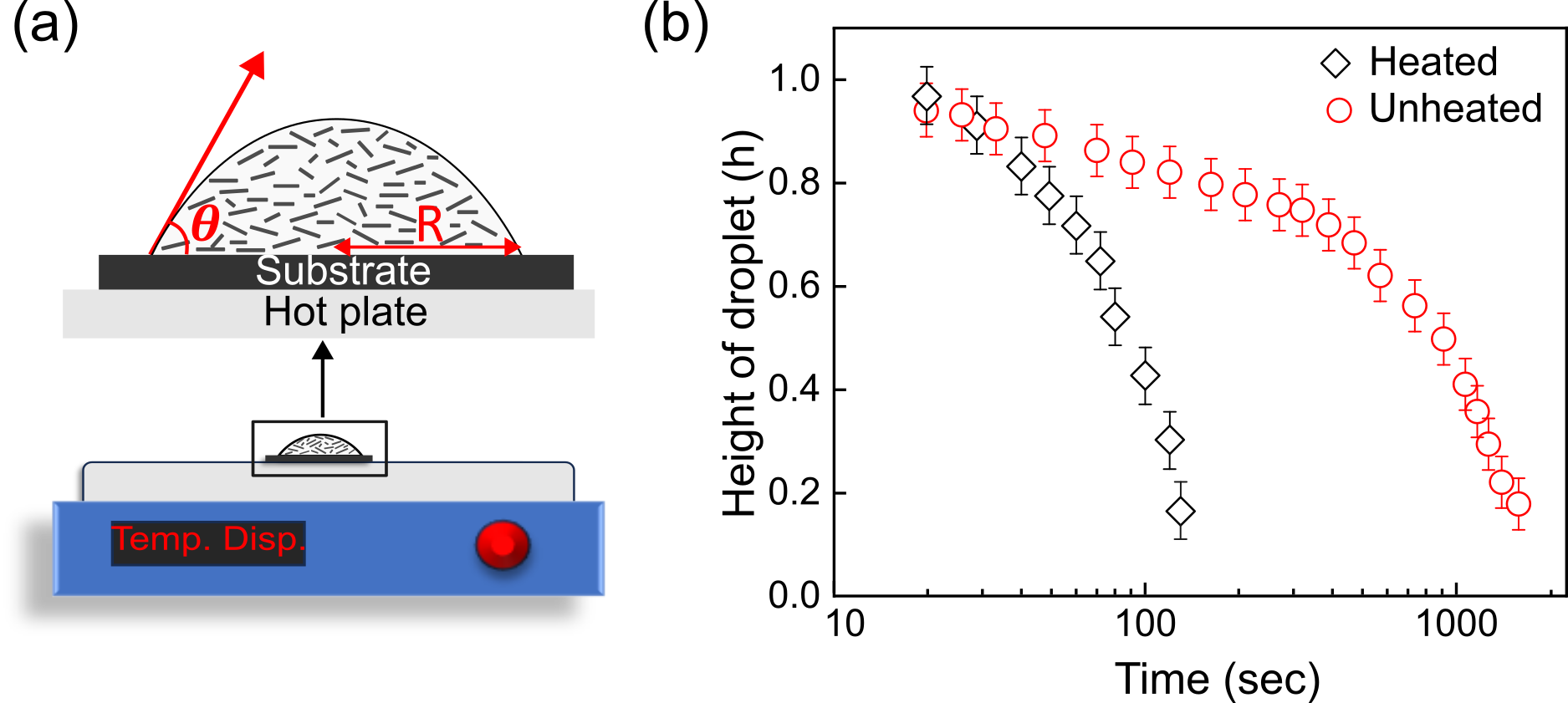}}
	\caption{Experimental protocol and evaporation dynamics: (a) Schematic of a sessile droplet containing HNTs placed on a heated substrate. The substrate is positioned at the centre of the hot plate and maintained at a desired temperature. $\theta$ and $R$ denote the contact angle and the droplet radius, respectively. (b) Droplet height ($h$) as a function of time (sec) for the unheated and heated cases. $h$ recedes $\approx 11$ times faster in heated case than the unheated case.}
	 \label{Figure-2}
   \end{figure}

To compare evaporation rates in the heated and unheated case, we monitor the droplet height ($h$), as a function of time. Figure~2b clearly indicates that substrate heating leads to a much quicker descent of the interface, implying a substantial enhancement in the evaporation rate.

\subsection{Height profile of the dried droplet}
Evaporation of sessile droplets containing colloidal suspensions typically undergoes the coffee-ring effect, where all the particles tend to move towards the edge \cite{deegan1997capillary}. However, our objective is to achieve a more uniform deposition upon evaporation. It is well-known that the heated evaporation suppresses the coffee-ring effect by driving solute particles away from the droplet edge during drying \cite{li2015coffee,parsa2015effect}. This effect has its origins in the emergence of the Marangoni flow, which is defined as the fluid motion induced by surface tension gradients that develop between the warmer edge and the cooler apex (room temperature) of the droplet \cite{lama2017tailoring} [See Figure 3a]. In contrast, the dominant fluid transport mechanism for the unheated case is the capillary flow which arises due to a non-uniform evaporation across the droplet surface \cite{lama2017tailoring}.

Although Marangoni flows are strongest at early stages when the temperature gradient is highest, they are known to persist throughout the drying process \cite{chatterjee2020evaporation}. Importantly, their magnitude far exceeds that of capillary flows (mm/s and $\mu$ms$^{-1}$, respectively), leading to a relatively more uniform redistribution of suspended particles during evaporation \cite{barmi2014convective,patil2016effects}. Insets in Figure 3b,c show representative photographs of the dried deposits for unheated and heated cases, respectively. They clearly show that in the heated case, the coffee-ring pattern is significantly thinner, leading to a more uniform spread of suspended particles across the substrate. This is also reflected in their corresponding height profiles, as shown in Figure~3b,c

\begin{figure}[t]
	\centerline{\includegraphics[width=0.48\textwidth]{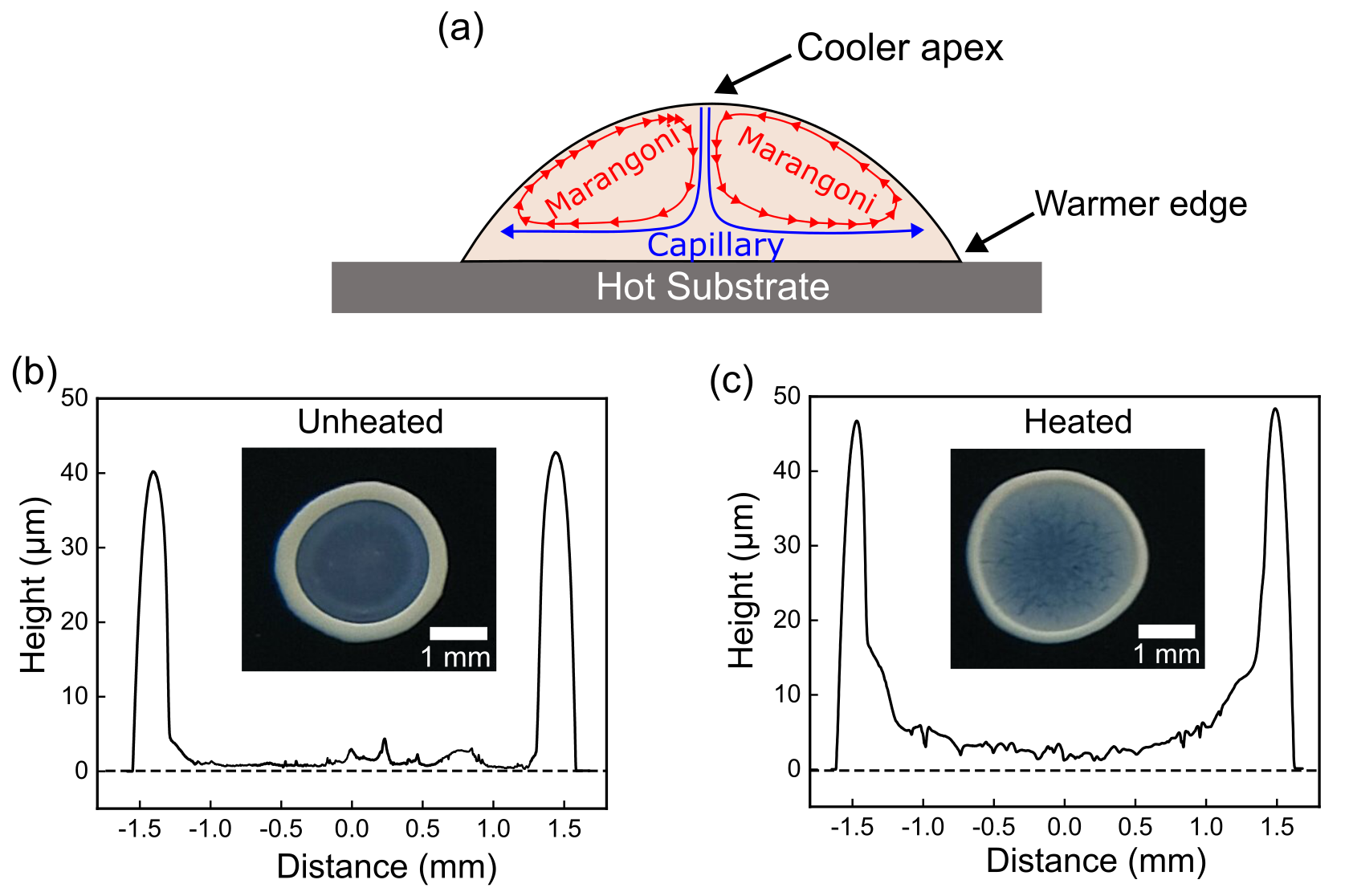}}
	\caption{Effect of Marangoni flow on the dried deposit and the height profile: (a) Schematic showing the underlying fluid flows (capillary and Marangoni) inside an evaporating sessile droplet. (b,c) Height profiles of the dried deposits under unheated and heated conditions, respectively. Insets show corresponding optical images of the dried deposits showing a more uniform distribution of suspended particles across the substrate in the heated case.}
	\label{Figure-3}
\end{figure}

\begin{figure*}
	\centerline{\includegraphics[width=1\textwidth]{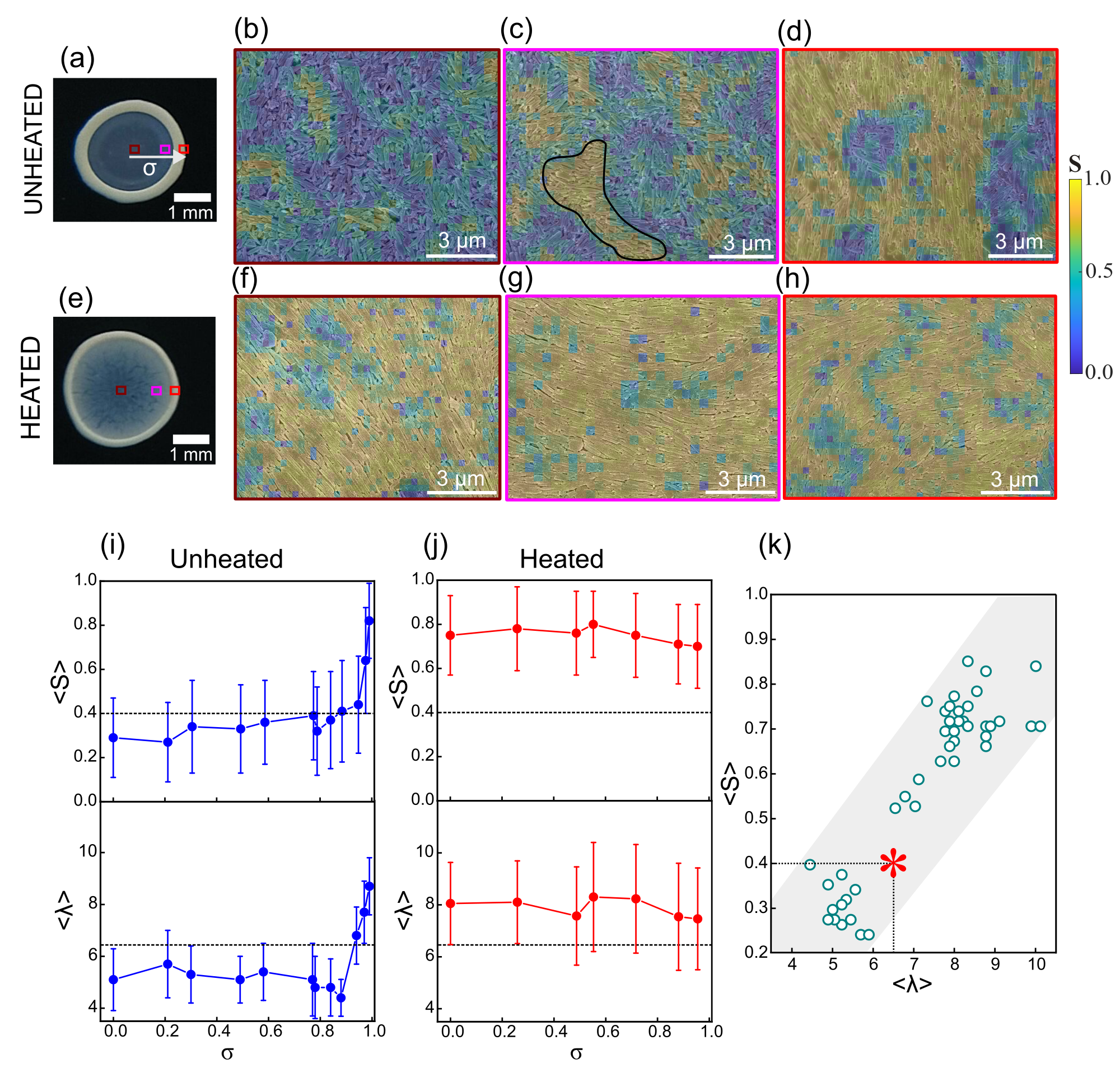}}
	\caption{Quantification of nematic order: (a) Dried deposit formed under the unheated condition, with corresponding SEM images (b-d) along the radial direction as a function of $\sigma$ (distance in units of droplet radius, $\sigma / R$) showing microstructural details at the marked positions in (a). Similarly, (e) shows the dried deposit formed under the heated condition, with corresponding SEM images (f-h) at marked positions. The color bar indicates the local nematic order parameter ($S$). (i-j) Variation of the $\langle S \rangle$ and $\langle \lambda \rangle$ as a function of the $\sigma$ under unheated and heated conditions, respectively showing similar trends. Error bars equal standard deviation. For the heated case in (j), large $\langle S \rangle$ and $\langle \lambda \rangle$ values occur concurrently. (k) Linear relationship between $\langle S \rangle$ and $\langle \lambda \rangle$ highlighted using the shaded region as a guide to the eye and asterix symbol highlighting the threshold values of $\langle S \rangle$ and $\langle \lambda \rangle$ of 0.4 and 6.5, respectively, signifying the onset of the nematic order in our system.}
	\label{Figure-4}
\end{figure*}

\subsection{SEM analysis and quantification of nematic order}
Having established the macroscopic uniformity of deposition under heated evaporation conditions, we next investigate the microscopic arrangement of the deposited HNTs using SEM. Our goal is to capture the full extent of the top surface morphology, from the center of the droplet to its outer edge. To systematically quantify the spatial variation, we define a normalized radial coordinate, $\sigma = r/R$, where $r$ is the radial distance from the center and $R$ is the droplet radius, which remains constant throughout the evaporation process. In our experiments, $R$ lies in the range of 1.4 \textendash 1.6 mm. Later, we take 10 \textendash 15 high-resolution SEM images at varying $\sigma$ values between 0 and 1.

\begin{figure*}
	\centerline{\includegraphics[width=1\textwidth]{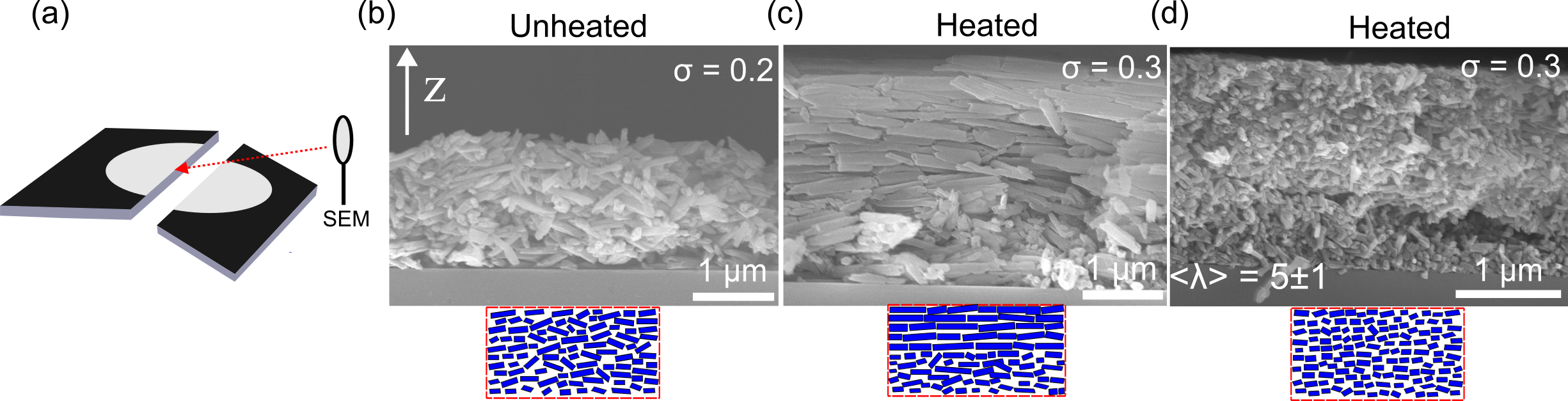}}
	\caption{Cross-sectional SEM imaging of dried deposits: (a) Schematic of the fractured silicon substrate used for SEM cross-sectional imaging. (b) SEM image of the unheated sample showing no indication of statification as well as orientation in HNTs. A schematic of the absence of stratification is shown below. (c) SEM image of the heated sample clearly showing stratification with longer rods occupying the top layers and showing a strong nematic alignment. A schematic of this result is shown below. (d) SEM image and a schematic of a sample with average aspect ratio $\langle \lambda \rangle = 5 \pm 1$, neither showing vertical stratification nor alignment.}
	\label{Figure-5}
\end{figure*}

Figures 4a,e show the dried deposits for the heated and unheated cases, respectively. Alongside morphological imaging, we also map the local nematic order using a superimposed color map on the SEM images. The 2D nematic order parameter is defined as $S = \langle2 \cos^2 \theta -1\rangle$, where $\theta$ is the angle between the rod's orientation and its local mean orientation computed over a square region of area 0.1 $\mu$m$^2$ \cite{cetera2014epithelial,dadwal2023quantifying} (See methods, Section 4.3). Clearly, by definition, $S$ = 0 and $S = 1$ corresponds to randomly oriented and perfectly aligned rods, respectively.

Figure~4b-d show that for the unheated case, $S$ remains low across most of the deposit, indicative of a disordered assembly. Near the coffee-ring edge, however, we observe localized patches of high $S$, previously referred to as \textit{nematic islands} \cite{dadwal2023quantifying}, consistent with a first-order isotropic–nematic transition (See Figure 4c). In stark contrast, for the heated case (Figure~4f-h), $S$ remains uniformly high across the entire deposit, indicating the presence of a global nematic order. This noticeable enhancement of the orientational order highlights the strong influence of substrate heating on evaporation-induced self-assembly of HNTs.

To gain deeper insights into the observed orientational order, we calculate the mean nematic order parameter, $\langle S \rangle$, by averaging the local $S$ values over each SEM micrograph's entire field of view over an area of $15\,\mu\text{m} \times 11\,\mu\text{m}$. Parallely, we also determine the average aspect ratio, $\langle \lambda \rangle$, for the rods within the same region. As shown in Figure 4i,j, a correlation emerges between $\langle S \rangle$ and $\langle \lambda \rangle$ when plotted as a function of $\sigma$ following similar trends in both unheated and heated experiments, respectively. Building on the findings of \cite{dadwal2023quantifying}, we adopt a threshold value of $\langle S \rangle = 0.4$ to differentiate between the isotropic and nematic phases, coinciding with $\langle \lambda \rangle$ value of $6.5$ to mark the onset of nematic ordering. Upon analyzing the $\langle S \rangle$ and $\langle \lambda \rangle$ data from our experiments, we find that the current experiments also produce similar threshold values, as shown in Figure 4. As a result, the data presented in Figure 4j shows that for the heated case, the system remains in a nematic phase which is a direct consequence of $\langle \lambda \rangle \ge 6.5$. However, such an order is missing in the unheated case (Figure 4i). These findings indicate that during the heated evaporation, longer rods get concentrated near the top surface where they collectively self-organize into an ordered nematic phase. 


\subsection{Cross-sectional SEM \texorpdfstring{$\&$}{\&} signatures of vertical stratification} 

A high degree of nematic order observed at the topmost layer suggests that during the drying process, longer rods with higher aspect ratios migrate toward the air-liquid interface. This naturally raises the question: where do the shorter rods go? To investigate this, we examine the cross-sectional morphology of the dried deposits using SEM, aiming to detect signatures of vertical stratification in our system. The dried film is carefully fractured, and the exposed cross-sections are imaged from the side to capture the layering along $z$-direction, perpendicular to the substrate, as shown in Figure 5.

Figures 5b,c show cross-sectional SEM images of the dried deposits for unheated and heated substrates, respectively. In the unheated case, rods of all lengths appear uniformly distributed along the $z$-direction, without any noticeable alignment or size-dependent segregation. This is also schematically depicted at the bottom of Figure 5b. In contrast, the heated case (Figure 5c) reveals a clear stratification: longer rods are aligned and concentrated near the top surface, while shorter rods occupy the bottom layers. This confirms that vertical stratification occurs for the heated evaporation case schematically shown at the bottom of Figure 5c. 

To further validate this observation, we prepare a sample with reduced polydispersity in $\lambda$ of (= 5 $\pm$1)  (See methods, Section 4.4), at a concentration of $c$ = 5 wt.$\%$. A 4 $\mu$L droplet is dried at $50^\circ C$ under identical conditions and cross-sectional SEM images are captured as shown in Figure 5d. Clearly, there is no evidence of stratification due to lower polydispersity along with the absence of nematic ordering, since $\lambda < 6.5$ in this case. A schematic illustration is shown at the bottom of Figure 5d.

\subsection{Nematic order weakens at elevated substrate temperatures}
\begin{figure*}[t]
	\centerline{\includegraphics[width=1\textwidth]{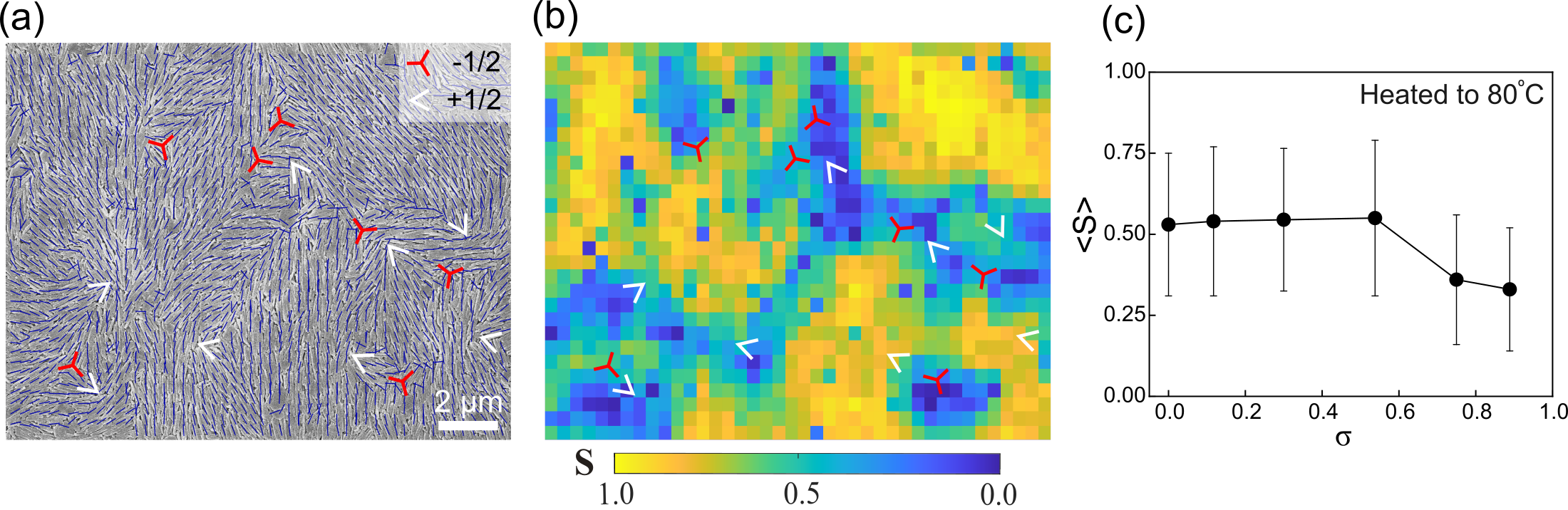}}
	\caption{Elevated substrate temperature reduces the nematic order: (a) SEM image of the HNTs with the local director field (blue lines) superimposed. Topological defects are indicated by red trefoil symbols for \( -1/2 \) defects and white chevron-shaped symbols for \( +1/2 \) defects. (b) Heat map of the local nematic order parameter, S, with the same topological defects marked as in (a). (c) Plot of $\langle S \rangle$ vs. $\sigma$, showing much lesser values of $\langle S \rangle$ at all $\sigma$ values at 80$^\circ$C. In comparison, $\langle S \rangle$ was much larger at 50$^\circ$C (Figure 4j). Error bars equal standard deviation.}
	\label{Figure-6}
\end{figure*}

To investigate whether similar phenomena are observed at even higher substrate temperatures, we conduct experiments with the substrate heated to 80$^\circ$C. Firstly, the evaporation rate increases threefold at 80$^\circ$C as compared to 50$^\circ$C, along expected lines. However, this also significantly alters the morphology of the top layer. A representative SEM image is shown in Figure 6a, overlaid with the nematic director field and locations of topological defects. Consistent with nematic symmetry, we observe only half-integer defects ($\pm 1/2$) \cite{de1993physics,selinger2016introduction,zhang2018interplay}, depicted by trefoil (red) and chevron-shaped (white) symbols for \textendash1/2 and +1/2, respectively. 

Clearly, these topological defects coincide with regions of low $S$ values (Figure 6b). As a result, in comparison to 50$^\circ$C case (Figure 4j), $\langle S \rangle$ is markedly reduced for all values of $\sigma$, as shown in Figure 6c. These findings imply that there exists an optimal substrate temperature for achieving uniform nematic order, and our results suggest this optimum lies close to 50$^\circ$C.

\section{Conclusions and discussion}

In this study, we presented an experimental protocol for achieving an in-plane nematic deposition of colloidal rods, starting from a polydisperse suspension of HNTs. The process includes a combination of temperature gradient-driven Marangoni flows and size-based vertical stratification, along with enhancement of orientational order beyond a critical aspect ratio of the rods. Our model system consisted of charge-stabilized HNTs, which rendered the rods effectively non-interacting. As a result, the system could be modeled using the framework of lyotropic liquid crystal physics. Through systematic SEM imaging, we demonstrated that the evaporation-induced stratification selectively traps longer rods at the air-water interface. Upon drying, these longer rods, with aspect ratios exceeding 6.5, occupy the top layers, exhibiting a strong nematic order. Cross-sectional SEM images confirm that the shorter rods occupy the lower layers of the dried deposit, supporting the stratification mechanism. In contrast, when the same experiment was conducted on an unheated substrate at room temperature, neither stratification nor nematic ordering was observed, highlighting the critical role of the thermal gradient. Interestingly, we found that increasing the substrate temperature to 80$^\circ$C beyond an optimal point (50$^\circ$C) leads to a reduction in nematic order due to the emergence of topological defects in the assembly. Nevertheless, the simple yet robust protocol described here may offer new possibilities for engineered surface coatings and functional materials without the requirement of homogeneous particle distributions.

It is worth noting that the emergence of vertical stratification in our experiments was an unexpected finding. According to conventional understanding, stratification results from the interplay between evaporation-driven advective flows and thermal diffusivity: the former promotes size-based sorting, while the latter tends to homogenize particle distribution. This is quantified through the \text{Pé}clet number, defined as $Pe = vH/D$, where $v$ is the liquid interface descent rate (proportional to evaporation rate), $H$ is the droplet height, and $D$ is the thermal diffusivity of the colloidal particles. At $T = 323\ \mathrm{K}$, we use the droplet height of \( H = 1\,\mathrm{mm} \) and the descent rate, $v$, estimated from the height-time plot (Figure 2b) to be $6.46 \times 10^{-6}\,\mathrm{m/s}$, respectively. Using the Stokes-Einstein relation $D = k_B T/6\pi \eta L$, where $L$ is the length of HNTs, and $\eta$ is the viscosity of water, the corresponding $Pe$ value for HNTs of length $0.4$ and $1.0\,\mu\mathrm{m}$ fall within the range of $10^3$ and $10^4$, respectively. This suggests that the thermal diffusion is negligible, which should prevent homogenization of the particles, thus preventing stratification. However, as noted by Jung et al. \cite{jung2020evaporation}, in the case of a dense suspension of rod-like particles, the relevant length scale $H$ may instead be the interparticle spacing, which is of the order of the rod length $L$. In this case, the $Pe$ in our system straddles around 1, reducing to 0.2 and 1.4 for $L = 0.4$ and $1.0\,\mu\mathrm{m}$, respectively, potentially explaining the observed stratification.

Another interesting observation in our experiments is the weakening of nematic order at elevated substrate temperatures, seemingly due to the emergence of topological defects. The precise mechanism causing these defects is difficult to pinpoint. One possible explanation is the presence of enhanced Marangoni flows at higher temperatures, which could generate chaotic interfacial dynamics that disrupt nematic ordering \cite{mur2022continuous}. Alternatively, rapid evaporation may quench the system into a prematurely arrested nematic phase, unable to fully develop, thus giving rise to topological defects. 

Finally, our experimental system based on HNTs offers a powerful platform to investigate the added complexity introduced by size polydispersity. The use of elongated particles allows us to leverage the well-established framework of liquid crystal physics, integrating it with the rich phenomenology of evaporation-driven self-assembly. While characterizing the final static structures is relatively straightforward, the real challenge lies in capturing the underlying dynamics, particularly how Marangoni flows mediate the transport and organization of irregularly shaped particles, a question that remains largely unresolved. Although technically challenging, real-time tracking experiments will offer a promising route to uncover how flow fields, particle shape, and interparticle interactions together govern structure formation in these complex, far-from-equilibrium systems. Success in this direction could lead to fundamental advances in our understanding of evaporation-driven ordering in complex, polydisperse colloidal suspensions.

\section*{Acknowledgements}
    NK acknowledges financial support from DST-SERB for CRG grant number CRG/2020/002925 and IITB for the seed grant. AD gratefully acknowledges IIT Bombay for providing a post-doctoral fellowship. AD thanks Prof. Sunita Srivastava for her valuable insights that contributed to this work and for permitting the use of her laboratory facilities. We thank SAIF-IITB for the FEG-SEM facility. AD thanks Somnath Paramanick and Bibaswan Ghosh for their assistance with the figures. 

\section{Methods}
\subsection{HNTs and their purification }
HNTs and SHMP were obtained from Sigma Aldrich, while sodium hydroxide (NaOH) was purchased from Merck Chemicals. HNTs were purified using the method reported in our previous paper\cite{dadwal2023quantifying}. In a typical procedure, 10 g of halloysite powder was added to a flask containing 100 mL of water, and the resulting mixture was sonicated for 20 minutes to disperse the particles. After sonication, the mixture was allowed to stand, and the settled portion was removed. The remaining suspension was then stirred continuously at 700 rpm. Under this constant stirring, SHMP (2 g) was gradually added to the above mixture, and after 15 minutes, the pH of the resulting mixture was adjusted to a range of 8\textendash9 using a 0.4 M aqueous NaOH solution. Subsequently, this mixture was stirred at room temperature for 24 hours and left standing for 12 hours to precipitate the aggregates. The impurities and larger HNTs aggregates precipitated at the flask's bottom while individual HNTs remained in the supernatant. The supernatant was collected and then centrifuged at 3200 rpm for 10 minutes. The resulting precipitates were washed twice using deionized water (DI) water. Finally, the obtained precipitate was dried at room temperature.

\subsection{Treatment of Si substrate} 
For evaporation studies, 1 cm × 1 cm single-side polished Si wafers of $\langle$100$\rangle$ orientation were used after cleaning with acetone and water. The substrates were first sonicated in acetone for 10 minutes, followed by sonication in DI for another 10 minutes, and then thoroughly dried before being used for the self-assembly studies.

The dried deposits of HNTs were investigated using a field emission gun (FEG)-scanning electron microscope (JSM 7600F) operating at an acceleration voltage of 10 keV. The zeta potential of the suspensions was measured using Zetasizer Nano-ZS analyzer.

\subsection{Image Analysis}
\textbf{Director field analysis:} We used ImageJ software~\cite{abramoff2004image} for initial image processing, including band-pass filtering to remove pixel noise and enhance brightness uniformity. Then, we applied the unsharp mask algorithm to sharpen rod edges. The processed images were analyzed using a 2D Fast Fourier Transform-based algorithm (FFT), following the method of Cetera et al.~\cite{cetera2014epithelial}, as previously described in our work~\cite{dadwal2023quantifying}. This algorithm performs a local 2D FFT on a small image local section ({area~$\mathcal A$}) to determine a vector orthogonal to the actual director field. Based on the SEM image resolution and clarity, the analyzed {area $\mathcal A$} typically ranges between $0.05$ and $0.2\,\mu\mathrm{m}^2$.

\subsection{Tuning the aspect ratio of HNTs}
This was achieved by slightly modifying the protocol reported in our previous paper\cite{dadwal2023quantifying}. First, the mixture was centrifuged at 3200 rpm to separate the supernatant, which was then subjected to subsequent centrifugation at 7000 rpm. After this step, the supernatant was passed through a 0.45 \textmu{m} nylon filter. The filtered solution was then centrifuged at 11,000 rpm. The resulting precipitate was dried at room temperature.

\bibliography{references}

\end{document}